\documentclass[11pt]{article}
\usepackage{amsmath}
\usepackage{cancel,soul,ulem}
\usepackage{pdfpages}
\usepackage{authblk}

\newcounter{continue}
\usepackage{times}
\usepackage{amsfonts}
\usepackage{graphicx}
\usepackage{wrapfig}
\usepackage{amsmath}
\usepackage{amssymb}
\usepackage{color}
\usepackage{cases}
\usepackage{subfigure}
\usepackage{tensor}
\usepackage{accents}
\usepackage[colorlinks=true, citecolor=blue]{hyperref}
\usepackage{xspace}
\usepackage{mathtools}
\usepackage{float}

\def\ben{\begin{enumerate}}
\def\een{\end{enumerate}}
\def\beq{\begin{equation}}
\def\eeq{\end{equation}}
\def\bea{\begin{eqnarray}}
\def\eea{\end{eqnarray}}
\def\bdoc{\begin{document}}
\def\edoc{\end{document}}
\def\nn{\nonumber}
\def\a{\alpha}
\def\b{\beta}
\def\k{\kappa}
\def\l{\lambda}
\def\la{\langle}
\def\ra{\rangle}
\def\up{\upp}
\def\down{\dn}
\def\e{\epsilon}
\def\d{\delta}
\def\D{\Delta}
\def\m{\mu}
\def\n{\nu}
\def\o{\omega}
\def\O{\Omega}
\def\r{\rho}
\def\s{\sigma}
\def\t{\tau}
\def\bfs{\mbox{\boldmath$\sigma$\unboldmath}}
\def\bfp{\mbox{\boldmath$p$\unboldmath}}
\def\bfA{\mbox{\boldmath$A$\unboldmath}}
\def\bfa{\mbox{\boldmath$\alpha$\unboldmath}}
\def\bfsp{\bfs\! \cdot\! \bfp}
\def\half{\textstyle{\frac{1}{2}}}
\def\fourth{\textstyle{\frac{1}{4}}}
\def\third{\textstyle{\frac{1}{3}}}
\def\3halfs{\textstyle{\frac{3}{2}}}
\def\em{\it}

\def\bfmu{\mbox{\boldmath$\mu$\unboldmath}}
\def\bfL{{\bf L}}
\def\bfS{{\bf S}}
\def\br{{\bf r}}
\def\bfp{\mbox{\boldmath$p$\unboldmath}}

\def\bA{{\bf A}}
\def\bD{{\bf D}}
\def\bH{{\bf H}}
\def\bM{{\bf M}}
\def\bN{{\bf N}}
\def\bE{{\bf E}}
\def\bF{{\bf F}}
\def\bB{{\bf B}}
\def\bP{{\bf P}}
\def\bJ{{\bf J}}
\def\bj{{\bf j}}
\def\bK{{\bf K}}
\def\bL{{\bf L}}
\def\bR{{\bf R}}
\def\bS{{\bf S}}
\def\bV{{\bf V}}
\def\bv{{\bf v}}
\def\bx{{\bf x}}
\def\by{{\bf y}}
\def\bz{{\bf z}}
\def\ba{{\bf a}}
\def\bd{{\bf d}}
\def\bi{{\bf i}}
\def\bj{{\bf j}}
\def\bn{{\bf n}}
\def\bm{{\bf m}}
\def\bp{{\bf p}}
\def\bk{{\bf k}}
\def\bg{{\bf g}}

\def\vare{\varepsilon}

\def\br{{\bf r}}
\def\rhat{\hat{\br}}
\def\bnab{{\bf \nabla}}

\def\bitP{\boldsymbol{P}}
\def\bphi{\boldsymbol{\phi}}
\def\btau{\boldsymbol{\tau}}
\def\bo{\boldsymbol{\omega}}
\def\bO{\boldsymbol{\Omega}}
\def\O{\Omega}

\def\tb{\tilde{\b}}
\def\tU{\tilde{U}}
\def\qd{\dot{q}}
\def\qdot{\dot{q}}
\def\xdot{\dot{x}}
\def\ydot{\dot{y}}
\def\thetadot{\dot{\theta}}
\def\phidot{\dot{\phi}}
\def\∂{\partial}
\def\w{\wedge}
\def\grad{\bnab}
\def\curl{\bnab\times}
\def\div{\bnab\cdot}

\def\uds{\underline{ds}}
\def\fdn{\underline{d\bn}\!\!\!\!\!\!\!\overline{\phantom{nn}}}
\def\fdv{\underline{dv_\m}\!\!\!\!\!\!\!\!\!\overline{\phantom{nn}}}
\def\fdvn{\underline{dv}\!\!\!\!\!\!\!\overline{\phantom{nn}}}
\def\fdR{\underline{d{\scriptstyle R}}\!\!\!\!\!\!\!\overline{\phantom{nn}}}

\newcommand{\be}{\begin{equation}}
\newcommand{\ee}{\end{equation}}

\newcommand{\lan}{\langle}
\newcommand{\ran}{\rangle}
\newcommand{\wt}{\widetilde}
\newcommand{\bs}{\boldsymbol}
\newcommand{\upp}{\uparrow}
\newcommand{\dn}{\downarrow}

\title{\Large Feynman 1947 letter on path integral for the Dirac equation
\vspace{1cm}}
\author{\large Ted Jacobson}
\affil{
\begin{small}
\it Maryland Center for Fundamental Physics\\
\it University of Maryland\\
\it College Park, MD 20742\\
\end{small}
}
\date{}
\begin{document}

\maketitle

\begin{abstract}
In 1947, four months before the famous Shelter Island conference, Richard Feynman wrote a 
lengthy letter to his former MIT classmate Theodore Welton, reporting on his efforts to 
develop a path integral describing the propagation of a Dirac particle. 
While these efforts never came to fruition, and were shortly 
abandoned in favor of a very different method of dealing with the electron propagator appearing in in QED, 
the letter is interesting both from the historical 
viewpoint of revealing what Feynman was thinking about during that
period just before the development of QED, and for its scientific ideas.
It also contains at the end some philosophical remarks, which Feynman wraps up with the 
comment, ``Well enough for the baloney.''
In this article, I present a transcription of the letter along with editorial notes, 
and a facsimile of the original
handwritten document. 
I also briefly comment on Feynman's efforts and discuss their relation to some later work.

\end{abstract}

\tableofcontents
\section{Introduction}

The motivating germ of Feynman's  diagrammatic formulation of QED 
was the idea that if the electromagnetic field is eliminated in favor of direct inter-particle interaction,
then it is possible---classically---to exclude the problematic self-interactions that lead to divergent self-energy \cite{Nobel, Schweber:1986pm}.
Wheeler and Feynman \cite{Wheeler:1945ps,Wheeler:1949hn}
developed a classical realization of this approach to electrodynamics
using a relativistic worldline action principle with pairwise 
interactions.\footnote{What was novel in their work was not the
action principle, but the recognition that if the universe is a perfect absorber of radiation 
then the physics of the time symmetric action principle would manifest as if one restricted to 
purely retarded (or purely advanced) electromagnetic fields, which could account for radiation
reaction.} 
To implement this idea in the quantum theory, it was necessary to develop a worldline 
action approach to quantum mechanics, which Feynman did for nonrelativistic particles in 
his Ph.D.\ dissertation \cite{Feynman:1942us}, and reported in a journal article six years later \cite{Feynman:1948ur}.
A next step in this program would have been to develop a relativistic worldline 
path integral for spin-1/2 particles, so that in particular electrons and positrons could be included in 
the theory. Feynman never published anything along these lines, but we know from 
comments in his Nobel Prize address \cite{Nobel}, and from a letter he wrote to his former MIT classmate
Theodore (Ted) Welton \cite{letter}, that this was not for want of trying.

The letter, which consists of thirteen densely handwritten pages,
came to light in Schweber's RMP article, {\it Feynman and the Visualization of 
Spacetime Processes}~\cite{Schweber:1986pm}, where it was mentioned briefly in Section IV.C.
In {\it The Genesis of Feynman Diagrams} by W\"uthrich~\cite{wuthrich2010genesis}, it was the subject of extensive discussion in Sections 4.3-4.5 of
Chapter 4, ``The Dirac Equation: Feynman's Great Struggle''. 
The focus there was on the role this effort played on the path to the diagrams, and on Feynman's motivations, his evaluation of his own
attempts, and on his attitude toward visualization and understanding in physics. These sections include transcriptions of roughly a page and a half of text from the letter, mostly pertaining to this focus, as well as facsimiles of all but four of the pages from the letter.
The present article, in contrast, focuses on parsing and evaluating 
the technical aspects of Feynman's efforts 
to construct a path integral for a Dirac particle in three spatial dimensions.

In view of historical interest in what Feynman was up 
to just before developing his diagrammatic approach to 
QED, as well as the fact that there has been a continuing 
small but persistent effort by others to develop a spacetime path integral for spin-1/2 particles
(without the use of Grassmann variables), it seemed  worthwhile to 
transcribe and reproduce the letter in full, to make it more accessible. 
My own interest was driven in large part by the fact that I had developed a similar
approach to the problem many years ago \cite{Jacobson:1983xt}, 
inspired by Problem 2-6 in the book by Feynman and Hibbs \cite{feynman2010quantum},
which lays out how the propagator for a 
Dirac particle in 1+1 spacetime dimensions can be 
obtained from a sum over paths that zigzag at the speed of light. 
It is fascinating to see in the letter how Feynman tried 
(without success as far as I can tell) to handle the 3+1 dimensional case. 

Section \ref{sec:letter} of this paper contains a complete transcription of the letter along with editorial footnotes aimed at clarifying its content.
Section \ref{note} contains a transcription of a brief note, probably written by Feynman, outlining a strategy for separating out the effect of the
mass term in constructing a path integral for the Dirac equation. 
Section~\ref{sec:commentary}~ presents commentary on, and evaluation of, Feynman's efforts, 
as well as discussion their relation to some later work.
Facsimiles of the original handwritten letter and note are included in Supplementary Material.

There is no finer way to close this introduction than to quote Feynman's own comments on his efforts in this
direction, taken from his Nobel Lecture \cite{Nobel}:
\begin{quote}
\it{
Another problem on which I struggled very hard, was to represent relativistic electrons with this new quantum mechanics. I wanted to do a unique and different way -- and not just by copying the operators of Dirac into some kind of an expression and using some kind of Dirac algebra instead of ordinary complex numbers. I was very much encouraged by the fact that in one space dimension, I did find a way of giving an amplitude to every path by limiting myself to paths, which only went back and forth at the speed of light. The amplitude was simple ($i\e$) to a power equal to the number of velocity reversals where I have divided the time into steps and I am allowed to reverse velocity only at such a time. This gives (as $\e$ approaches zero) Dirac's equation in two dimensions -- one dimension of space and one of time ($\hbar=m=c=1$).

Dirac's wave function has four components in four dimensions, but in this case, it has only two components and this rule for the amplitude of a path automatically generates the need for two components. Because if this is the formula for the amplitudes of path, it will not do you any good to know the total amplitude of all paths, which come into a given point to find the amplitude to reach the next point. This is because for the next time, if it came in from the right, there is no new factor $i\e$ if it goes out to the right, whereas, if it came in from the left there was a new factor $i\e$. So, to continue this same information forward to the next moment, it was not sufficient information to know the total amplitude to arrive, but you had to know the amplitude to arrive from the right and the amplitude to arrive to the left, independently. If you did, however, you could then compute both of those again independently and thus you had to carry two amplitudes to form a differential equation (first order in time).

And, so I dreamed that if I were clever, I would find a formula for the amplitude of a path that was beautiful and simple for three dimensions of space and one of time, which would be equivalent to the Dirac equation, and for which the four components, matrices, and all those other mathematical funny things would come out as a simple consequence -- I have never succeeded in that either. But, I did want to mention some of the unsuccessful things on which I spent almost as much effort, as on the things that did work.
}
\end{quote}
\section{The letter}
\label{sec:letter}

The letter \cite{letter} is filed at the Caltech Archives in Box 11, Folder 2 (Dirac Equation). 
It responds to a prior letter from Welton, which is likely the 18 page letter that 
is in the same Folder 
(except for the first two pages which are missing).
Many other papers and notes related to the Dirac equation are filed in Box 11, Folder 2,
although I found no significant development on a path integral for the 3+1 dimensional Dirac equation beyond what appears in this letter.

I have transcribed the letter preserving in most cases notation, punctuation, and 
placement of text, except when that 
would have been impractical due to the placement of diagrams, side notes, or annotations
on the page. Crossed out text has been omitted, 
and the equation numbers are added by me in the transcription
for ease of future reference.
Anything in square brackets is clarification or commentary by me, 
including all of the footnotes.
A facsimile of the letter is included in the Supplementary Material.\\
\hspace{1cm}
\noindent\rule{\textwidth}{1pt}

\begin{flushright}
Monday,  Feb.\ 10.\footnote{[Schweber's bibliography in \cite{Schweber:1986pm} cites 1947 as the year of this letter,
and W\"uthrich \cite{wuthrich2010genesis} points out that February 10 fell on a Monday only in 1941 and 1947 during the 1940's.]}
\end{flushright}

Dear Ted

I got your very interesting letter today.

I realized how 2nd quantization in the case of a He atom transformed something that looked like Schrodinger's picture (if $\psi$ is a function) to something that was correct (if $\psi$ is an operator). So I knew you are on the right track with your theory \& I now believe it to such an extent that I will learn 2nd quant.\ \& try to solve your problem of how to deal with the ``rest of the universe."\footnote{[Welton's letter is in part concerned with the effort to develop a relativistic quantum field theoretic version of the Wheeler--Feynman 
absorber theory of radiation, based on half advanced plus half retarded potentials.]}
 I am engaged now in a general program of study -- I want to understand (not just in a mathematical way) the ideas in all branches of theor.\ physics. As you know I am now struggling with the Dirac Eqn.

The ideas which you outlined were, as you say, vague, but otherwise they are exactly right in the sense that the way I now see the Dirac Eqn.\ is just a more precise formulation of your (\& of course my earlier) ideas. There are several ways I can explain it, all leading to the same answer but all not equally precise. The one with the most shaky physical ideas at present, I think is the closest to the truth \& I would like to explain it to you because (1) you may get nice ideas (2) I will learn something by formulating it clearly. In this particular ``model" paths do not necessarily go at light velocity -- altho I may change that later as you shall see. 

The idea which I got to in 2 dimensions, of loading each turn thru $\theta$ by $e^{i\theta/2}$ does not work well in 3 dimensions for the angles $\theta$ are in different planes. What is needed is to keep track of 3 dimensional rotations, and to use a factor instead of $e^{i\theta/2}$ which is capable of keeping track of a complete 3-dimensional rotation. 
%%
%\begin{wrapfigure}{o}{0.08\textwidth}
%  \begin{center}
%    \includegraphics[width=0.07\textwidth]{Fig.1}
%  \end{center}
%%  \caption{}
%\end{wrapfigure}
%%
To be more precise, one model I had said this: go on any path at light velocity but consider only paths that start from A \& go to B entering B in the same direction as they left A.  
\begin{figure}[h]
\centering\hspace{1cm}
    \includegraphics[width=1.5cm]{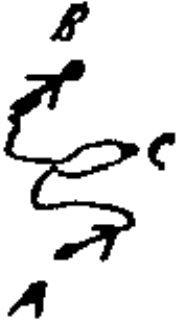}\hfill
      \includegraphics[width=3cm]{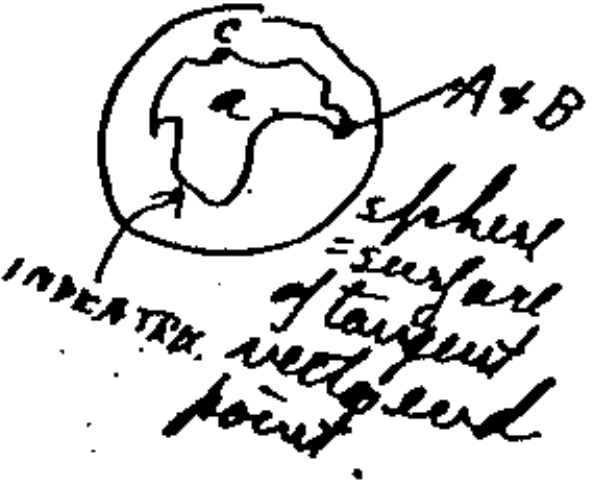}\hfill
        \includegraphics[width=3cm]{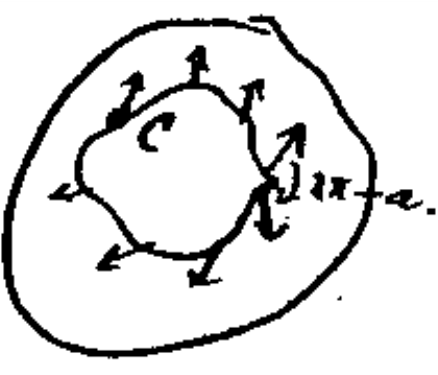}\hspace{1cm}
\end{figure}
%
%\begin{figure}
%     \centering
%     \begin{subfigure}{0.3\textwidth}
%         \centering
%         \includegraphics[width=\textwidth]{Fig.1}
%        % \caption{$y=x$}
%         %\label{fig:y equals x}
%     \end{subfigure}
%     \hfill
%     \begin{subfigure}{0.3\textwidth}
%         \centering
%         \includegraphics[width=\textwidth]{Fig.2}
%        % \caption{$y=3\sin x$}
%         %\label{fig:three sin x}
%     \end{subfigure}
%     \hfill
%     \begin{subfigure}{0.3\textwidth}
%         \centering
%         \includegraphics[width=\textwidth]{Fig.3}
%        % \caption{$y=5/x$}
%         %\label{fig:five over x}
%     \end{subfigure}
%        %\caption{Three simple graphs}
%        %\label{fig:three graphs}
%\end{figure}
Then the factor is $e^{ia/2}$ where $a$ is the solid angle enclosed by the tangent (velocity) vector in its gyrations along the path. Now we at once ask what information must be kept by the time we reach C. 
At C the area (solid angle) 
%%
%\begin{wrapfigure}{R}{0.18\textwidth}
%  \begin{center}
%    \includegraphics[width=0.17\textwidth]{Fig.2}
%  \end{center}
%%  \caption{}
%\end{wrapfigure}
%%
enclosed by the vector is not closed so we have simply to keep track, sort of, of where we are on the sphere. But slightly more than that. When we get back to A (at B)\footnote{[This presumably refers to the directions at A and B.]} we have enclosed area \& that is what we have to keep track of. One geometrical way we can keep track is to put up a small arrowhead at A  \& move it parallel to itself as you go along the trajectory (differential parallelism). When it gets to B (I'm on the sphere surface now) it will have rotated, relative to its original position by an angle $2\pi - a$.
%%
%\begin{wrapfigure}{R}{0.18\textwidth}
%  \begin{center}
%    \includegraphics[width=0.17\textwidth]{Fig.3}
%  \end{center}
%%  \caption{}
%\end{wrapfigure}
%%
Thus at C we need keep only the axis of C and a direction for our little vector. This we can do by choosing a standard direction arbitrarily at each pt of the sphere and saying that at C we have to rotate around C by a given angle to define the angle of the little vector with the standard direction. Thus we have at the point C in the path to keep track of a \underline{rotation} in three space (around the axis of the tangent to the path). Thus the ``wave function" (defined to be that quantity or quantities which if known at C from \underline{all} paths reaching C would permit calculation [of] the same quantities at the next instant of time) must be a symbol representing a rotation in space -- which can be added \& which, if the rotation is ``$a$" around some particular axis (say the z axis) must be represented by $e^{ia/2}$.
[Added as a footnote:] I forgot to say, the whole thing can be put in a  very succinct way: any path represents a rotation. For imagine that we have a sphere with a pole \& keep turning it so that the pole-center line lies in the direction of the path's tangent. Then when we complete the path the sphere has made a net rotation. If the wave function at $A$ is $\psi_A$, that at $B$ represents the ``rotation of the path'' $\times \psi_A$ (summed on all paths). If tangent at $A$ is in same direction as at $B$, net rotation can only be rotation around pole-center axis. It is, by angle $a$.

I don't expect you to understand the above argument because it was a long time for me to figure it out too. However, be that as it may, when I studied quaternions which I knew were designed to represent rotations I realized that they were the mathematical tool in which to represent my thoughts. To remind you:\footnote{\label{quat}[Here 
in the right margin Feynman adds the properties of the basis quaternions:  $\bi=$ rot by 180${}^\circ$ 
about $x$, $\bj=$ rot by 180${}^\circ$ 
about $y$, $\bk=$ rot by 180${}^\circ$ about $z$, as well as the multiplication rules, $\bi^2 = -1$, $\bj^2 = -1$, $\bk^2 = -1$, $\bi\bj=\bk = -\bj\bi$ etc, and $q = \a + A_x\bi + A_y \bj + A_z\bk$. Near the end of the letter (see footnote \ref{Pauli2}) 
he mentions the relation to Pauli matrices (e.g.\ $\bi = - i \s_z$), referring to a notation that Welton had used.]} 
 a quaternion $q$ is a hybrid addition between a scalar $\a$ \& vector $\bA$.\footnote{[Feynman writes a vector as well as the corresponding quaternion as a letter with a double line on the left. In this transcript instead boldface notation is employed.]}
$q = \a + \bA$ say.\footnote{\label{Pauli}[For modern readers, it may be easier to follow the reasoning
by identifying $q=\a + \bA$ with the $2\times2$ complex
matrix $\a I -i\bA\cdot\bfs$, where $\bfs$ is the vector of Pauli matrices.]}
 If $q = \a + \bA$ , $p = \b + \bB$ then $qp = \a\b -\bA\cdot\bB + \a \bB + \b \bA + \bA\times\bB$.
$q+p = \a+\b + \bA+\bB$. Symbol $\bar q$ is defined (say adjoint quaternion $\leftarrow$ my term) $\bar q = \a - \bA$. The size of a quaternion squared is $q\bar q= (\a + \bA)(\a-\bA) = \a^2 + \bA\cdot\bA$: a scalar. The quaternion representing a rotation about the axis $\bn$ (unit vector) by angle $\theta$ is $\cos\theta/2+\bn \sin\theta/2 = e^{\bn\theta/2} = q_{\theta,\bn}$.\footnote{[With the Pauli matrix interpretation of footnote \ref{Pauli}, 
this is the unitary matrix $\exp(-i\theta\bn\cdot\bfs/2)$, 
which is the spin-1/2 representation of a rotation through angle $\theta\bn$ 
acting on two-component spinors. 
Alternatively, the conjugation $q_{\theta,\bn}\bA\bar q_{\theta,\bn}$ 
yields the quaternion corresponding to the rotated vector $\bA$.]} 
Such a quaternion has unit size. If $\bn$ is a unit vector, then considering it as a quaternion it represents a rotation by 180$^{\circ}$ around $\bn$. $\bn\bn = -\bn\cdot\bn + \bn\times\bn = -1$ so rotation by 360$^\circ$ around any axis is represented by $-1$. 720$^\circ$ by $+1$ \& is equivalent to no rotation. If $\bn_1$ and $\bn_2$ are two unit vectors, what quaternion is $-\bn_1\bn_2 = \bn_1\bar \bn_2$? It is the result of 180$^\circ$ around $\bn_2$ followed by -180$^\circ$ around $\bn_1$. This, by an argument due to Wigner which is clear when you think about it, is a rotation around an axis $\perp$ to $\bn_1$ \& $\bn_2$ by an angle equal to \underline{twice} the angle between $\bn_1$ \& $\bn_2$ (because 180$^\circ$ around $\bn_1$ moves $\bn_2$ to the other side, thus moves it by twice the angle between $\bn_1$ \& $\bn_2$.\footnote{[In the Pauli matrix representation of footnote \ref{Pauli},  \[\bn_1\bar \bn_2\longleftrightarrow
(-i\bn_1\cdot\bfs)(i\bn_2\cdot\bfs)= \bn_1\cdot\bn_2 + i  (\bn_1\times\bn_2)\cdot\bfs = \cos\theta_{12}I + i\sin\theta_{12}
\bn_{12}\cdot\bfs = \exp(i\theta_{12}\bn_{12}\cdot \bfs),\] 
where $\theta_{12}$ is the angle of rotation 
from $\theta_1$ to $\theta_2$ and $\bn_{12}$ is the unit vector in the direction of $\bn_1\times\bn_2$. 
This seems to differ by the sign of the rotation from what Feynman says, but it is not entirely clear to me what he meant.]}

Consider then a path, which may be looked upon as a sequence $\bn_1$, $\bn_2$, $\bn_3,\dots$ of unit vectors 
(tangent vectors) (velocity vectors) at successive positions of the path. Suppose there are an even number of them (I'll do that better later). We seek the quaternion representing the rotation for this path resulting from turning $\bn_1$ to $\bn_2$, from $\bn_2$ to $\bn_3$, from $\bn_3$ to $\bn_4$ \dots etc. This is needed (see footnote above) for quantum mech. I assert it is simply $\bn_1\bar\bn_2\bn_3\bar\bn_4\dots$ (product of all $\bn$'s).\footnote{[It seems to me that the order of
multiplication should be the opposite.]} $\bar\bn = - \bn$. For consider that $\bn_1$ \& $\bn_2$ differ only infinitesimally (?). Then $\bn_1\bn_2$ is a rotation by \underline{twice} the angle $\bn_1$ to $\bn_2$, \& then $\bn_3$ to $\bn_4$ is a rot.\ by twice angle $3\rightarrow4$. But angle 2 to 3 or 4 to 5 etc are left out, so assuming continuity (?) etc.\ the factor of two is undone! (What I say is OK in the limit for a \underline{continuous} path -- I don't know what is true for Weierstrass curves with infinite discontinuities in slope or something). 

Thus the factor $\Phi = e^{\frac{i}{\hbar}S}=e^{\frac{i}{\hbar}\int_{t_1}^{t_2} L(x,\dot x) \, dt}$
in my non-relativistic quantum mech.\ must be replaced by $\bn_1\bar\bn_2\bn_3\dots\bar\bn_N$, 
and the wave function is a quaternion.\footnote{[Feynman does not comment on 
how he would connect a quaternion wave function to the spinor wave function that is governed by the Dirac equation.
Perhaps it was just obvious to him that he would eventually take the quaternion to operate on an initial 
two-component spinor, thus yielding another spinor.]} 
Thus one might guess, by reducing the long product to a final step, 
\underline{\underline{not}} $\psi(\br, t+\e) = \int \bn \psi(\br -\e\bn,t) \fdn$ but rather
\begin{align}
\psi(\br, t+\e) &= \int \bn \chi(\br -\e\bn,t) \fdn\label{psi1}\\
\chi(\br, t+\e) &= -\int \bn \psi(\br -\e\bn,t) \fdn
\end{align}
because with an odd no.\ of $\bn$ factors one has quite a different thing (called $\chi$)  than one has with even no. The symbol $\fdn$ means some sort (?)(!) of an average over all directions. The right hand side can be simplified, if we assume $\fdn$ means just average over all directions of the vector $\bn$; 
\begin{align}
{\cal E}g:\;
\int \bn \chi(\br - \e\bn,t)\fdn &= \left(\int \bn \fdn\right)\chi(\br,t) - \e\left(\int \bn(\bn\cdot\nabla)\fdn\right)\chi(\br,t) + \cdots\\
\bigl(\mbox{Now } \int\bn\fdn = 0,\quad &\int \bn(\bn\cdot\bA)\fdn = \bA\cdot\tfrac{1}{3}\bigr)
\label{Now}\\
&=0 - \tfrac{1}{3}\e\nabla\chi \qquad\qquad \label{O(e)}
\end{align}
here $\nabla\chi$ is a quaternion combination, not just $\nabla$ of a scalar.\footnote{\label{fail}[The implication of \eqref{O(e)} is that 
$\psi(\br, t+\e)$ in \eqref{psi1} is smaller than $\chi(\br, t)$ 
by a factor of $O(\e)$. This shows that the amplitude rule proposed in the
previous paragraph cannot yield the Dirac propagator, since every step would
entail another factor of $\e$. Feynman's comparison with \eqref{Dirac} 
suggests that he may have intended to subtract 
$\psi(\br, t)$ from the left hand side of \eqref{psi1}, 
in which case that side would also be $O(\e)$. However,
the resulting equation would not correspond to his proposed amplitude rule.]} 
It is to be figured thus:  if $\chi = \a + \bA$, then 
\beq
\nabla \chi = \nabla(\a + \bA) =  \nabla\a + \nabla\bA = (\nabla\a)  -(\nabla\cdot\bA)+(\nabla\times\bA).
\eeq
Now, the factor of 3 and a few other things worry me about this but I realize I am close because in quaternion notation Dirac's equations (with mass) (no field) can be written:
\begin{align}\label{Dirac}
\begin{split}
\frac{\partial a}{\partial t} &= \nabla b + i\mu b\\
\frac{\partial b}{\partial t} &= -\nabla a + i\mu a
\end{split}
\end{align}
so it looks like we are close to the equations with mass $\mu = \frac{mc}{\hbar}$ equal to zero.\footnote{[I suspect
that in the first equation $\nabla b$ should be $\nabla a$, and in 
the second equation vice versa, presuming that $a$ and $b$ are supposed to describe opposite chiralities.]}
Or maybe if we average just right, or add another effect we could get mass. (Note the imaginary.) To get the fields, multiply 
$\bn_1\bar\bn_2\bn_3\dots$ by $\exp \frac{ie}{\hbar c}\int(\bn\cdot\bA) ds$\footnote{[Annotations indicate that the integral is ``along path", and the integrand quantity in parentheses is ``ordinary scalar".]}

\noindent\hrulefill

Now, if you are still staggering but on your feet I will tell you more, which is even harder to understand.
My purpose now is to consider a path as a set of four functions $x_\mu(s)=x,y,z,t(s)$ of a parameter $s$. Thus the speed need not be that of light --- but for simplicity I want to restrict it to velocities less than that (or at most equal to $c$). The idea now is to consider a path is a succession of 4-vectors $v_\mu^{(1)}$, $v_\mu^{(2)}$, $v_\mu^{(3)}\dots$ representing successive proper velocities ($v^\mu v_\mu = 1$) on the path. Then each path represents a net Lorentz-Rotation transformation. (I mean a combination of a Rot.\ \& a Lorentz). Thus take the Lorentz trans.\ which carries 
$v_\mu^{(1)}$ to $v_\mu^{(2)}$. Then that which carries $v_\mu^{(2)}$ to $v_\mu^{(3)}$ etc. What is the net? We desire a symbolism, which by analogy works in 4 space (3+1) like the quaternions do in 3. I shall show that this symbolism is furnished by quaternions using complex numbers for components!

I have trouble visualizing 3+1 space, but I have become an expert at visualizing (really!) 4 dimensions with all dimensions equivalent. $ds^2 = dx^2 + dy^2 + dz^2 + dt^2$. What are the properties of rotations in this space? First a simple rotation called turn from on vector $\bn_1$ to another $\bn_2$ is not around just one axis, but around \underline {two}, (since there are 2 perpend.\ directions both $\perp$ to $\bn_1$ \& $\bn_2$). More precisely for any plane (e.g.\ of $\bn_1$ \& $\bn_2$) there is a completely $\perp$ plane (I call it the antipodal plane) such that if $\bn_1$ is any vector in the first plane, \& $\bn_2$ is \underline{any} vector in the second plane, then $\bn_1$ is $\perp$ to $\bn_2$. A plane (\& hence its antipodal plane) is determined by 4 parameters \{2 perpendicular unit vectors = 8 - 2 (unit) - 1(perp) = 5 nos.\ but they can be set at any angle in the plane $\therefore$ 4 para.\}. A complete 4 dim.\ rot.\ is characterized by 6 parameter \& can always be resolved into a turn (1 parameter) in a certain plane (4 para.) and in addition some other rotational angle (turn) in the antipodal plane (1 para). Turns in two antipodal planes commute, obviously. (Example a Lorentz trans.\  from velocity in $z$ direction is a turn in the $zt$ plane, this commutes with space rotations about the $z$ axis (turns in the $xy$ plane).).

Now (lets call these octonions\footnote{[This is not the standard meaning
of the term ``octonion''. Below, Feynman will conclude that, in Lorentz signature,
what he calls octonions can be identified with quaternions with complex
coefficients.]}) the ``quaternions'' representing 4 space rotations will probably have among their fundamental unit vectors, a set of six: $I_{xy}$, $I_{yz}$, $I_{zx}$, $I_{xw}$, $I_{yw}$, $I_{zw}$ such that $I_{xy}^2 = -1$ etc. and such relations as are true with one dimension lacking are still true, such as $I_{xy}I_{yz} = I_{zx} = -I_{yz}I_{xy}$. (We can identify $I_{xy} = \bk$, $I_{yz} = \bi$, $I_{zx} = \bj$). These fundamental units represent turns of 180${}^\circ$ in the various planes $xy$ etx. Hypothesis; (I) the octonion representing a rotation of $\varphi$ in the $xy$ plane is 
\beq
e^{\frac{\varphi}{2}I_{xy}} = \cos \frac{\varphi}{2} + \sin\frac{\varphi}{2} I_{xy}.
\eeq
Etc.\ for other planes. (II) If we go $\varphi$ on $xy$ plane, and $\theta$ on antipodal $zw$ plane the octonion is\footnote{[For the last term Feynman wrote  
$\sin\frac{\varphi}{2} \cos \frac{\varphi}{2}I_{xy}I_{zw}$, but this must have been a slip of the pen, since the expression should be symmetric under $\varphi\leftrightarrow\theta$ interchange. Also, obviously if $\theta=0$ there should be no term involving $I_{zw}$.]}
\beq
e^{\frac{\varphi}{2}I_{xy}} e^{\frac{\theta}{2}I_{zw}} =
 \cos \frac{\varphi}{2} \cos \frac{\theta}{2} + \sin\frac{\varphi}{2} \cos \frac{\theta}{2}I_{xy}
 + \sin\frac{\theta}{2} \cos \frac{\varphi}{2}I_{zw} + \sin\frac{\varphi}{2} \sin \frac{\theta}{2}I_{xy}I_{zw}.
\eeq
There is no ambiguity about the order of factors since rotations about the antipodal planes commute.
(hence also $I_{xy}I_{zw} = I_{zw}I_{xy}$).

I have just explained how to define the octonions for a special pair $xy$ \& $zw$ but it is clear what to do for any two antipodal planes (the $I$ for any plane is the appropriate (by coordinate transformation as tensor 2${}^{\rm nd}$ rank) linear combination of the $I$'s for $xy$, $xz$, etc.\ - In fact define $I_{yx}=-I_{xy}$ \& the $I$'s form \& transform like an antisym.\ tensor 2${}^{\rm nd}$ rank). 

Now we need one other fact to finish the description. We see we have new ``unit vectors'' or quantities such as $I_{xy}I_{zw}$. How many of these are there (for other pairs of planes e.g.\  $I_{xz}I_{yw}$) \& what are their properties? Well they represent two 180${}^\circ$ turns, one around each of two antipodal planes. The first turns $x$ to $-x$ and $y$ to $-y$; the second $z$ to $-z$ and $w$ to $-w$. Hence, (III) two 180${}^\circ$ rot.\ about antipodal planes, represented hereafter by the symbol ${\cal R}$ simply reverses direction of all axes and \underline{is the same no matter which pair of antipodal planes is used}.
Further, clearly ${\cal R}$ commutes with every operation. In 4 space ${\cal R}^2=1$ (since  ${\cal R}^2=360^\circ$ around two axes, \& each one is $-1$.  As I find, but don't \underline{see}, ${\cal R}^2 = -1$ in (3+1) dimens.\ space. Hence 
 \beq
 I_{xy}I_{zw}=-{\cal R} =I_{yz}I_{xw} 
 \eeq
 etc.\footnote{[Pointing to the minus sign in front of ${\cal R}$, Feynman inserts ``arbitrary choice for convenience''.]}
 Hence multiplying both sides by $I_{xy}$, note:
 \beq
 I_{xy}I_{xy}I_{zw}=-{\cal R}I_{xy} =-I_{zw}.\qquad\quad \therefore \boxed{I_{zw}={\cal R}I_{xy}} 
 \eeq
 hence we have 4 unit vectors , $\bi$, $\bj$, $\bk$, ${\cal R}$.  ${\cal R}$ commutes with all!
 \beq
 {\cal R}^2 =
 \begin{array}{rl}+1 & \mbox{(4 space)}\\1 & \mbox{(3+1 space)}\end{array}
 \qquad \bi^2= \bj^2=\bk^2=-1.\quad \bi\bj=\bk=-\bj\bi.\quad {\cal R}\bi = \bi {\cal R}\;\; \mbox{etc.}
 \eeq
 \beq
 \begin{array} {ll}
 I_{xy}=\bk & \qquad I_{xw}={\cal R}\bi\\
 I_{yz}=\bi &  \qquad I_{yw}={\cal R}\bj\\
 I_{zx}=\bj & \qquad  I_{xw}={\cal R}\bk\\
 \end{array}
 \eeq
 Every octonion then is of the form, 
 \beq
 q = \a_1 + {\cal R}\a_2 + \bA_1 + {\cal R}\bA_2
 \eeq
 where $\a_1$ and $\a_2$ are nos.\ and $\bA_1$, $\bA_2$ represent 3 vectors, $\bA_1=A_{1x}\bi + A_{1y}\bj + A_{1z}\bk$, 
 $\bA_2=A_{2x}\bi + A_{2y}\bj + A_{2z}\bk$. Thus, an octonion is a quaternion whose coefficients are of form $a +{\cal R}b$. In 3+1 space
 ${\cal R}^2=-1$ and we can take ${\cal R}=i$ if we wish, so octonions for Lorentz Trans \& Rotations are quaternions with complex $(a+ib)$ coefficients. QED. The result for 4 space is independent of which axis is used for $w$, but in 3+1 space, to preserve the 3+1 symmetry we can take the $t$ axis (in some Lorentz frame) to be $w$.
 
 Now, how can we represent the transformation needed to turn $v_\m$ into $u_\m$ where $v_\m$, $u_\m$ are unit four vectors? (In 4 space, first). Consider the octonion $+(\sum_\m v_\m u^\m) - \sum_{\m>\n} v_\m u^\n I_{\m\n}$.\footnote{[Feynman probably meant to write the contravariant form $v^\m$, but at many points he does not seem to be concerned with index placement.]} I say it represents a turn in the $v$, $u$ plane by \underline{twice} the angle of the turn of $v_\m$ to $u_\m$. (Because, the analysis above showed us we could take any axis, so choose $x$, $y$ in plane of $v$, $u$. $z$ \& $w$ perp. Then $v_\m u^\m=\cos\theta$ where $\theta$ is angle between and 
  $\sum_{\m>\n} v_\m u^\n I_{\m\n}=(v_x u_y - u_x v_y)\bk = \sin\theta\,\bk$.\footnote{[Feynman  indicated above $\bk$ that it is equal to $I_{xy}$.
(Also, he mistakenly wrote $u_y v_x$ for the second term.)]} $\therefore q = \cos\theta + \bk \sin\theta = e^{\theta \bk}$ while the turn 
  $v_\m$ to $u_\m$ is $e^{\theta \bk/2}$). Problem is, can we represent it as a product $vu$ of a quaternion depending on $v$ only by one depending on $u$ only. Yes. Choose any arbitrary axis (say $w$). Then consider two quaternions $v_\m$ into $w_\m$ by twice angle, and $w_\m$ into $u_\m$ by twice angle. Then $v_\m$ into $u_\m$ by twice angle = result of $v_\m$ into $w_\m$ by twice $\times$  $w_\m$ into $u_\m$ by twice. (Proof: Find the axis perpend. to $u_\m$, $v_\m$, $w_\m$. Consider all rotations that keep this axis fixed. Then we are dealing with just 3-space rotations \& can use Wigner's idea that $v_\m$ into $u_\m$ by twice angle = 180$^\circ$ around $v_\m\times 180^\circ$ around $u_\m$. Hence 
  \[
  v_\m\rightarrow u_\m\, \mbox{by 2 angle} = v_\m\rightarrow w_\m\, \mbox{by 2 angle} \times w_\m\rightarrow u_\m\, \mbox{by 2 angle}
   \]
   becomes
 \[
 180^\circ\mbox{ on }v_m\times -180^\circ \mbox{ on }u_m =180^\circ\mbox{ on }v_m\times -180^\circ \times w_m\times 180^\circ\mbox{ on }w_m\times -180^\circ \mbox{ on }u_m
 \]
 but the two $180^\circ$ on $w_\m$ cancel out proving the lemma.)
 \[
 \therefore\; \sum_\m(v_\m u^\m)-\sum_{\m>\n}v_\m u^\n I_{\m\n} = v_\m\rightarrow u_\m\mbox{ at 2 $\times$ angle } = v\bar u
 \]
 where $v = v_\m\rightarrow w_\m\mbox{ at 2 $\times$ angle } =  \sum_\m(v_\m w^\m)-\sum_{\m>\n}v_\m w^\n I_{\m\n}$. 
 $u$ is defined likewise. $\bar u$ means $u$ with every $\bi$, $\bj$, $\bk$ (but \underline{NOT} ${\cal R}$) reversed in sign. 
 If $w_\m$ is chosen for convenience as the 4$^{\rm th}$ axis,
 \beq\label{A}
 v = v_4 - (v_1\bi + v_2 \bj + v_3 \bk){\cal R} \hspace{2cm}\tag{A}
 \eeq
 where ${\cal R}=i$ in 3+1 dimensions.\\
 
 Hence, the Lorentz transformation defined by a path is 
 \[
 v^{(1)}\bar v^{(2)}v^{(3)}\bar v^{(4)}\dots
 \]
 where $v_\m^{(i)}$ is the $i^{\rm th}$ velocity (proper velocity) vector in succession and $v^{(i)}$ is the quaternion 
  $v_4^{(i)} - i(v_1^{(i)}\bi + v_2^{(i)} \bj + v_3^{(i)} \bk)$. (A)\\
  
 Hence the wave equations resulting from the obvious analogy from 3 dim.\ case are ($s$ is a parameter on wave function, $R_\mu$ means four vector position, time).\footnote{[Feynman indicated ``Quat. (see eqn (A))" above $v$ in the first integrand.]}
  \beq\tag{B}\label{B}
  \begin{aligned}
  \psi(R_\m,s+\e) &= \int v\chi(R_\m+\e v_\m,s) \fdv\,. \\
  \chi(R_\m,s+\e) &= \int \bar v\psi(R_\m+\e v_\m,s)\fdv \,.
  \end{aligned}
  \eeq
 Here $\fdv$ is an integral in an invariant way over all directions (in forward light cone) equally weighted. Assume $\psi$, $\chi$ independent of $s$, or varying exponentially with $s$ and you obtain Dirac's eqn.\ with mass by inserting appropriate constants (for fields add factor $e^{i\sum_\m\int v_\m A^\m ds}$ whose significance I am now pursuing). Or again do not assume $v_\m$ is of unit length but write $\int v \chi(R_\m + v_\m; s) F(\sum v_\m v^\m) dv_1 dv_2 dv_3 dv_4$ where $F$ is an arbitrary function. Mass is determined by the eigenvalue (values?) of $F$. Or again, take most simply for $F$,\footnote{[Pointing to the prefactor $i$, Feynman writes ``Meaning what?"]}
  $i\d'(v^\m v_\m)$. That is, 
 \beq\label{C}\tag{C}
 \psi(R_\m) =ic\int(R-R')\chi(R_\m')dR_1dR_2dR_3dR_4\;\d'\bigl((R_\m - R_\m')^2\bigr)
 \eeq
  where $R$ is a quaternion $R_4 - i(R_1\bi + R_2\bj + R_3\bk)$ and choose $c$ appropriately. This goes back to the light cone idea and everything is back, but I point out the 4 dimens.\ view because it lends (?) more complete understanding.
  With this choice for $F$, you see we get exactly your equations because\footnote{\label{Pauli2}[Pointing to $(R-R')$  after $\therefore$ Feynman writes ``quat'', and after that equation he adds parenthetically ``(Needless to say, what you call $\sigma_x$ I call $i\bi$; $\sigma_y = i \bj$, $\s_z = i\bk$.)"]}
  \beq
  (R_\m - R_\m') \d'\bigl((R-R')^2\bigr)\equiv \frac{\partial}{\partial x_\m}\d\bigl((R-R')^2\bigr)
  \eeq
  \beq
  \therefore (R-R')\d' \equiv \Bigl\{\frac{\partial}{\partial t} - i \Bigl(\bi \frac{\partial}{\partial x} +\bj \frac{\partial}{\partial y} +\bk \frac{\partial}{\partial z}\Bigr) \Bigr\} \d\bigl((R-R')^2)\bigr)
  \eeq
  So 
  \begin{align}
  ic \int (R-R')\,\chi(R_\m')\,\d'\,\fdR &=   ic \int  \Bigl(\frac{\partial}{\partial t} -i\bnab\Bigr)\,\d\,\chi\;\fdR \nn\\
  &= ic\Bigl(\frac{\partial}{\partial t} -i\bnab\Bigr)(\square^2 \chi) \nn\\
  &= ic\,k^2\Bigl(\frac{\partial}{\partial t} -i\bnab\Bigr)\chi \nn\\
  &=\psi\,.\label{delta'integral}
  \end{align}

  Now I would like to add a little hooey. The reason I am so slow is not that I do not know what the correct equations, in integral or differential form are (Dirac tells me) but rather that I would like to \underline{understand} these equations from a many points of view as possible. So I do it in 1,2,3,\& 4 dimensions with different assumptions etc. It is just that vagueness which you mention which I would like to try to clear up. I was especially impressed by your intuition in guessing just what the $\psi$'s would mean (I mean the odd \& even deflections, etc.). I hope you get something out of my point of view. I am attempting this to see how to alter things in a natural way. (Eg.\ should the $\d$ function be exactly a $\d$ function?) What is interaction between particles? For fields it is easy to define as $e^{iv\cdot\bA}$ factor, which is sensible if the $v$ is infinitesimal per step -- and in the integrals on the left of \eqref{B} and \eqref{C} ought to extend over short distances so $e^{iv\cdot\bA}$ is OK, but right along light cones the distances get long -- finite in fact, but usually one gets no contribution from large distances because \underline{usually} the $\psi$ or $\chi$ function oscillates rapidly, but I want to investigate this more thoroughly. 
  It is very interesting that if you integrate $\frac{e}{\hbar c}\int A_\m dx^\m$ along a long finite section of straight light ray the biggest contributions come from electrons exactly in the direction of the ray and the result is then simply
  $\frac{e^2}{\hbar c} \ln(r_1/r_2)$, where $r_1$ and $r_2$ are the distances (or times) of the end pts of integration away from the electron producing the field.\footnote{[Feynman seems to have assumed here that the ray and the inertial charge wordline lie in a common plane, and he has chosen Coulomb gauge for the vector potential (the integral in question is not gauge invariant).]}
  Can all electrodynamics be stated by this simple logarithm rule?
  
 I find physics is a wonderful subject. We know so very much and then subsume it into so very few equations that we can say we know very little (except these equations - Eg.\ Dirac, Maxwell, Schrod). Then we think we have \underline{the} physical picture with which to interpret the equations. But there are so very few equations that I have found that many physical pictures can give the same equations. So I am spending my time in study -- in seeing how may new viewpoints I can take of what is known.
 
 Of course, the hope is that a slight modification of one of the pictures will straighten out some of the present troubles.
 
 I dislike all this talk of there not being a picture possible, but we only need know how to go about calculating any phenomena. True we only \underline{need} calculate. But a picture is certainly a \underline{convenience} \& one is not doing anything wrong in making one up. It may prove to be entirely haywire while the equations are nearly right -- yet for a while it helps. The power of mathematics is terrifying -- and too many physicists finding they have the correct equations without understanding them have been so terrified they give up trying to understand them. I want to go back \& try to understand them. What do I mean by understanding? Nothing deep or accurate --- just to be able to see some of the qualitative consequences of the equations by some method other than solving them in detail.
 
 For example, I'm beginning to get a mild ``understanding'' of the place of Dirac's $\a$ matrices which were invented by him ``to produce an equation of first order in the differential coefficient in the time'', but by me in order ``to keep track of the result of a succession of changes of coordinate system''. Still my stuff sounds mathematical -- \& insofar as it is, I still don't understand it --- but I will try soon to reformulate in terms of seeing how things look to someone riding with the electron. (For instance when an electron appears to go completely around a proton, the proton does not appear from the point of view of the electron to complete its circuit around the electron! (See Thomson\footnote{[Feynman probably intended to refer to Thomas (precession) here.]} on spin orbit interaction.) Because the succession of Lorentz transformations come out at the end as a small rotation of the electron's axes).
 
 Why should the fundamental laws of Nature be so that one cannot explain them to a high-school student -- but only to a quite advanced graduate student in physics? And we claim they are simple! In what sense are they simple? Because we can write them in one line. But it takes 8 years of college education to understand the symbols. Is there any simple \underline{ideas} in the laws?
 
 Well enough for the baloney. I'll get back to work on understanding better my new Dirac Equ. (By the way I did all this before I got your letter, except to notice that the choice leading from \eqref{B} to \eqref{C} gave the Dirac Eqn.\ in a particularly direct way --- but I did notice that any function $F(v_\m v^\m)$ gives Dirac equations if correctly normalized. I was engaged in studying this: Certainly
 \beq\label{certainly}
 \int v\,\chi(R_\m-v_\m)\, \fdvn\;F(v_\m v^\m) = 
 \int v\,\Bigl(v_\m \frac{\partial}{\partial x_\m}\chi\Bigr)\, \fdvn\,F(v_\m v^\m) =
 \Bigl\{\Bigl(\frac{\partial}{\partial t} -\bnab\Bigr)\chi\Bigr\} \underbrace{\int v_\m v^\m \,F(v_\m v^\m) \,\fdvn}_\text{a number}
  \eeq
 so Dirac's Eqn.\ comes out. But if $v_\m v^\m$ is nearly zero (assuming $F(s^2)$ is small unless $s^2$ is small) it might still be true that $v_\m$ is not small enough to allow replacement in $\chi(R_\m - v_\m)$ of $\chi$ by its first order Taylor expansion $\chi(R) + v_\m \frac{\partial}{\partial x_\m}\chi$. 
 
 Thanx for your suggestions and I look forward to hearing anything else you might think about this. If you don't want to read this letter over \& over could you send it back so if someone else wants an explanation they can read it.
 
 I will work on your 2$^\text{nd}$ quantization, when I get time. I really believe it is correct -- but is another example of the terrifying power of math.\ to make us say things which we don't understand but are true. \\
 
 \hspace{3in} Sincerely
 
 \hspace{3.9in} Dick\,.
 
 \noindent\rule{\textwidth}{1pt}
 
 \section{Note on strategy for separating the effect of the mass}
 \label{note}
 
This section contains a transcription of a note, filed 
in Box 11, Folder 2, in which the author, who is probably Feynman, 
sets out a strategy for separating out the 
 effects of the mass term in seeking a path integral for the Dirac equation. 
My guess is that the note was written after the letter to Welton, since the strategy  did not seem to be clear to Feynman in the letter.
 A facsimile of the original note
 is included in the Supplementary Material.
 
 %\hspace{1cm}
 \noindent\rule{\textwidth}{1pt}
%\hspace{1cm}
 %
 \begin{enumerate}
 \item If a rule of track will work then if it will apply for very small time intervals $\e$ then the mass term in the Dirac Equ. can be neglected ($\e\ll\hbar/mc$). Hence look at neutrino. 
 \item Field effects are also small.
 \item The is no length scale for (free) neutrino case.
 \item Hence either free neutrino rule is exact.
 \item Or it is not readily reduced to infinitesimal $\e$. 
 \end{enumerate}
 Hence either find exact neutrino rule (Huygens Princ.) or expect a rule that shows up correctly only after an 
unlimited number of steps!\\

\noindent Thus all you need is a rule that works OK when $\psi$ is nearly constant -- i.e.: \underline{small} gradients!

 \noindent\rule{\textwidth}{1pt}
 
 \section{Commentary and relation to later work}
\label{sec:commentary}
\begin{itemize}

\item First of all, it is important to emphasize that the letter is an informal,
private communication with a friend, describing work in progress, so should not
be held to the standards one expects of a publication or even of an
informal report on established results.

\item Feynman was strongly motivated to ``understand'' the Dirac equation from an elementary viewpoint that could be guessed from simple principles, as he says both in the letter and in his Nobel Lecture quoted in the Introduction. 
Rather than just working backward from a finite-differenced 
Dirac or Weyl equation, which would have been easier, 
Feynman aimed to infer the path integral on purely  geometric 
and algebraic grounds, using the fact that the spin is correlated
with the displacement of the particle, together with 
relations between rotations and quaternions. It appears that
his motivation for that came in part from the case in two spatial dimensions, 
which he mentions in the letter. 
The notes in the archives contain some analysis of the two dimensional case.

\item The role that Feynman's unsuccessful attempts at
a path integral for the Dirac particle played in his progress toward
the fantastically successful perturbative diagrammatic method 
for QED is explored in depth in Ref.~\cite{wuthrich2010genesis}. There are
at least two facets to this role. One is the mere fact that, 
in formulating quantum mechanics in terms of a path integral,
Feynman was led to draw spacetime diagrams associated to 
complex numbers that are summed to obtain the 
full quantum transition amplitude. This carries over
directly into Feynman diagrams. Another facet is that,
lacking a path integral for 
the free Dirac particle transition amplitude, Feynman 
shifted attention to the object that this path integral 
should have yielded, namely, the propagator, which ends up attached 
to a diagram line that replaces 
the role of a spacetime path.
His experience associating quaternions to path steps would have made the introduction of 
the Dirac-matrix-valued propagator a natural construction.

\item Feynman's intuitive, heuristic methods 
that failed for the Dirac path integral were 
nevertheless effective in navigating 
his way from the classical Wheeler-Feynman formulation of 
electrodynamics and the Dirac equation to his diagrammatic, 
perturbative formulation of quantum field theory.  
Illuminating accounts and analyses of Feynman's 
development of these ideas
can be found in Refs.~\cite{Nobel, Schweber:1986pm, wuthrich2010genesis},
as well as in more recent articles 
by Blum~\cite{Blum:2017diy} 
and by Darrigol~\cite{Darrigol:2019qlz}.

 \item Feynman's description of the path amplitude defined in terms of the solid angle on the sphere of tangent directions is not entirely clear to me,
 but it seems to be very close to Polyakov's spin factor \cite{Polyakov:1988md}. 
 Some problematic aspects of Polyakov's proposal are pointed out and corrected in \cite{Jacobson:1988yq},
 by working backwards from the retarded propagator
 for a (massless) Weyl spinor, 
expressing it as a path integral over a sequence 
 of steps defined by spinors. 
 The result appears to agree with what Feynman seems to be saying, at least concerning the
 phase associated with a path.
 Feynman's ``small arrowhead'' sounds like the ``flag'' of Penrose's 
``flagpole plus flag'' representation of a two component spinor (up to a sign) \cite{Penrose:1985bww}. 
The wave function at a point would thus correspond, 
apart from the sign ambiguity, to a such a spinor, as in \cite{Jacobson:1988yq}.
However, in \cite{Jacobson:1988yq} it is argued that a more careful
consideration of the continuum
limit indicates that the correct step speed is not 
the speed of light (as both Feynman and Polyakov said) but rather three times that.
That the discrete step speed should exceed the speed of light is consistent with 
the Courant--Friedrichs--Lewy step speed convergence condition,
which is required 
for a finite differenced hyperbolic equation to converge
to the continuum equation.
 
 \item   Although, as noted in footnote \ref{fail},
 Feynman's quaternion chain guess for the 
 spin-1/2 path integral is not correct as it stands,
 it is quite close to the 
 approach taken in \cite{Jacobson:1983xt} which does work. 
The method in \cite{Jacobson:1983xt}
is to first derive a path integral for the chiral (massless) case,
 and then to describe a Dirac particle of mass $m$ by a pair of spinors
 of opposite chirality, which at each step of time duration $\e$ 
have an amplitude $i\e m$ to flip chirality.
To derive the chiral path integral, 
the partial derivatives in the 
Weyl equation are replaced by finite differences,
and the value of the two-component spinor wave function $\psi(\bx,t+\e)$
is expressed in terms of the values of $\psi(\bx',t)$ 
at the points $\bx'$ on a sphere of radius $3\e$ (as it turns out) 
surrounding $\bx$. The amplitude  for the step from $\bx'$ to $\bx$  
is obtained by multiplying the spinor 
$\psi(\bx',t)$ by the spin projection operator
$\half(I \pm \bn\cdot\bfs) = |{\pm}\bn\ra\la{\pm}\bn|$, 
where the $\pm$ sign corresponds to the fixed chirality of the Weyl particle,
$\bn$ is the unit vector pointing from $\bx'$ to $\bx$, and 
$|{\pm}\bn\ra$ is the eigenspinor of positive spin in the ${\pm}\bn$ direction
($\bn\cdot\bfs|{\pm}\bn\ra=\pm|{\pm}\bn\ra$).
The spin projection operator is very close to Feynman's quaternion path step, 
but he was missing the scalar part ($\half I$) of the quaternion 
that corresponds to the spin projection operator. 
Iteration of this one-step amplitude evolution yields a path integral 
in which a path is determined by a sequence of unit vectors $\bn_i$, and the path amplitude is
the product of the corresponding spin projection operators. 
The result is a path integral for the retarded 
propagator of a spinor wave function
in which the steps move at 3 times the speed of light, 
and which converges to the continuum propagator. 

\item The second approach Feynman sketches uses complex quaternions
(which are also known as biquaternions, whose algebra is equivalent to that
of the complex $2\times2$ matrices). The aim is a  
Lorentz covariant formulation, involving path steps that are 4-vectors. After developing 
some relations between the algebra of biquaternions and four-dimensional rotations or Lorentz transformations, he describes the proposed amplitude for a path, 
and then writes down a one-step evolution rule in Eqn.\ \eqref{B}.
A number of aspects look problematic. I may have misunderstood; but, for what it's worth,
my critiques are the following:
\begin{itemize}
\item The argument of the wave function $\psi(R_\m,s)$ is the four spacetime coordinates  
$R_\m$ and a path parameter $s$. The usual Dirac wave function is a function on spacetime, and the role of $s$ is unexplained.
\item The integral is over a Lorentz-invariant measure on 
the unit future timelike 4-vectors. Since a unit timelike vector can be arbitrarily ``long'', the proposed amplitude evolution rule is highly nonlocal in spacetime.
\item It is not clear  why the step vectors are chosen  {\it future} pointing rather than past pointing.
\item The unit future  timelike 4-vectors form a noncompact hyperbolic space, with infinite volume, so the integral in \eqref{B} is generally not convergent. 
\item Feynman goes on to consider other measures for the integral, one
of which is the derivative of a delta function of $v^\m v_\m$. 
With this measure he sees an indication
of the Dirac equation emerging in \eqref{delta'integral},
however I have been unable to follow the equality in the second line. 
(In the final line he seems to assume 
that $\chi$ has the form of a plane wave with squared wave 4-vector $k^2$, and has dropped a minus sign.)
\item The first equality of \eqref{certainly} seems to be missing 
a contribution from the zeroth order term in the Taylor expansion of $\chi$.
Such a term was not present in the prior construction with quaternions, but in \eqref{certainly} the biquaternion $v$ contains a scalar piece $v^0 I$ which does not integrate to zero.
\item As Feynman recognizes below \eqref{certainly}, the truncation of the Taylor expansion
is not justified even when $v^\m$ is null or nearly null, because the components 
of such $v^\m$ can nevertheless be ``large'' (in some fixed coordinate system).  
\item The second equality of \eqref{certainly} relies on the fact that 
the integral of $v^\m v^\n$ with respect to a Lorentz-invariant measure 
is proportional to the (inverse) Minkowski metric $\eta^{\m\n}$.
While this tensor structure is formally correct, 
for any choice of the weight $F(v^\m v_\m)$ other than 
zero the integral is not convergent. 
\end{itemize}

\item A brief review of work on 
path integrals for particles with spin
in the framework of  discrete time quantum walks, beginning with Feynman's 
checkerboard \cite{feynman2010quantum}, 
 can be found in section i.3.1 of 
\cite{Arnault:2017vae}.

\end{itemize}

 \section*{Acknowledgments}
I am grateful to the Feynman Estate for permission to reproduce the letter, for which it holds the copyright.
This work was supported in part by NSF grant PHY-2309634
and by Perimeter Institute where part of this work was carried out. Research at Perimeter Institute is supported in part by the Government of Canada through the Department of Innovation, Science and Economic Development and by the Province of Ontario through the Ministry of Colleges and Universities.

%\newpage

\bibliographystyle{unsrt}

%\enlargethispage{10\baselineskip}

\bibliography{Feynman_letter}

%\newpage

\appendix
\section{Supplementary material: facsimiles}
\includepdf[pages=-,pagecommand={},width=1.3\textwidth]{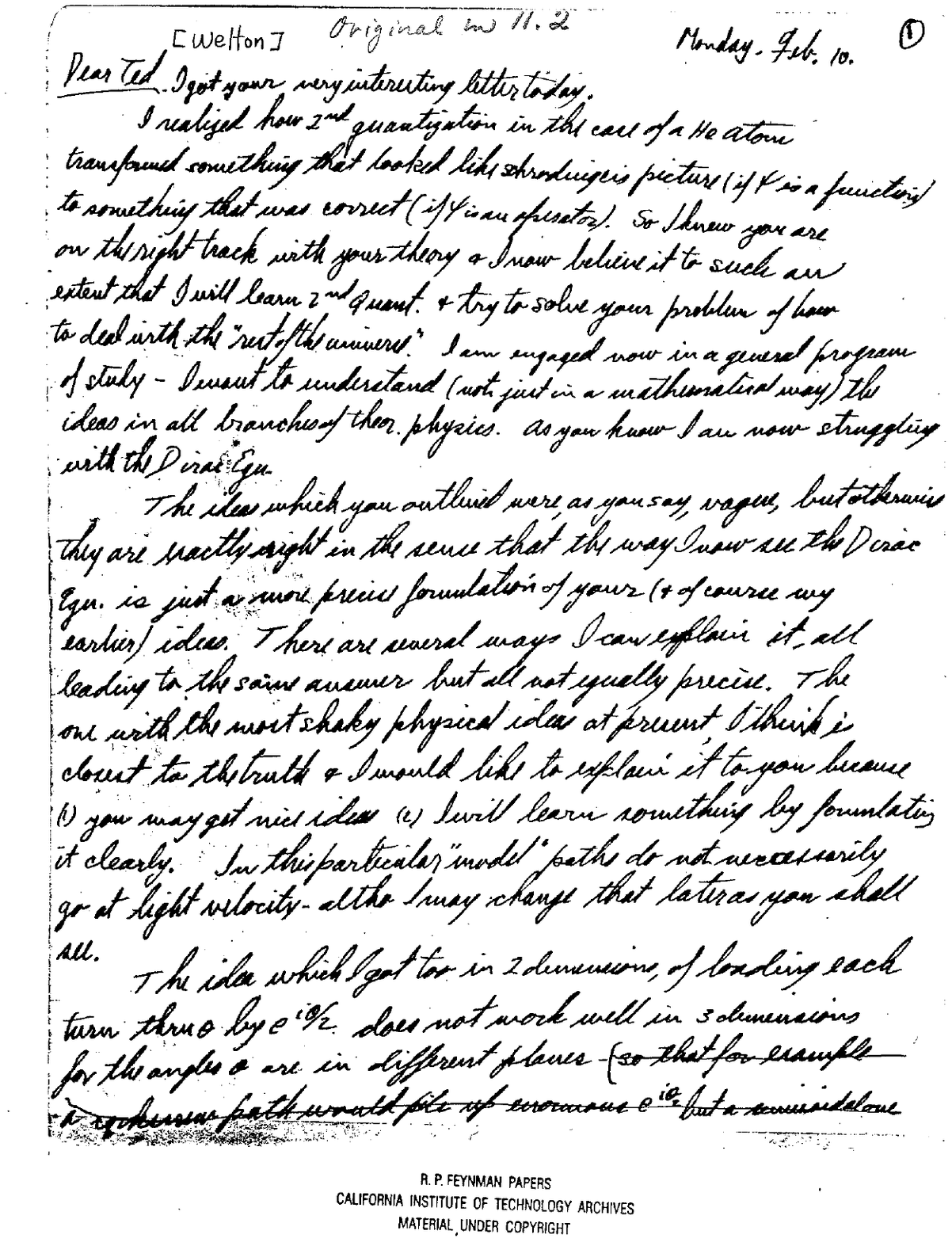}

\includepdf[pages=-,pagecommand={},width=1.3\textwidth]{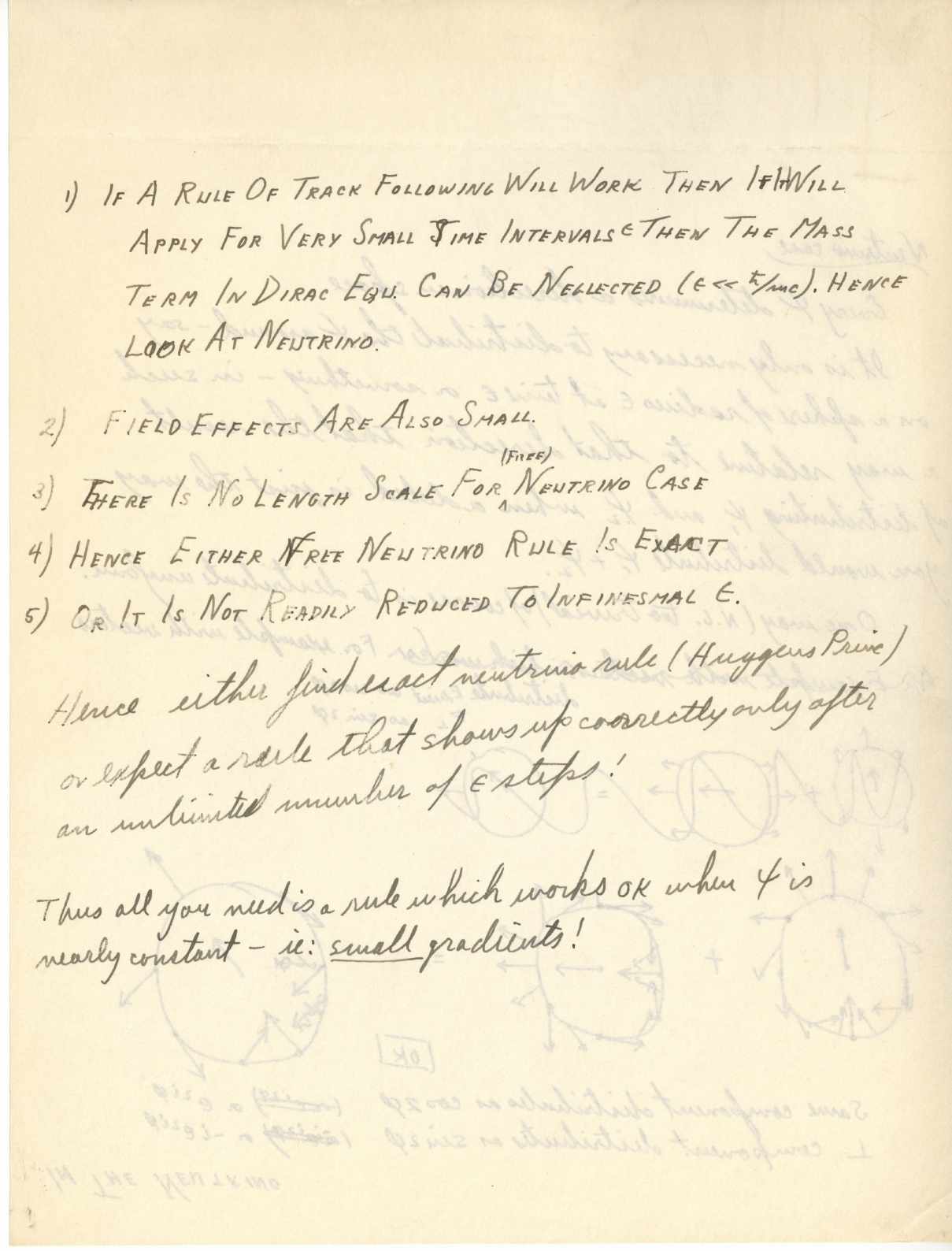}

  \edoc